\documentclass[prd,aps,showpacs,superscriptaddress,preprint]{revtex4}

\usepackage{graphicx}
\usepackage{amsmath}                             

\newcommand{\be}{\begin{equation}}
\newcommand{\ee}{\end{equation}}
\newcommand{\beq}{\begin{eqnarray}}
\newcommand{\eeq}{\end{eqnarray}}
\newcommand{\bea}{\begin{eqnarray}}
\newcommand{\eea}{\end{eqnarray}}
\def\eq#1{eq.~(\ref{#1})}

\newcommand{\tr}{\operatorname{tr}}
\newcommand{\re}{\operatorname{re}}

\begin{document}

\title{Meson masses and decay constants from unquenched lattice QCD.}

\preprint{DESY 09-095, LTH 833, SFB/CPP-09-56, HU-EP-09/28}

\author{K. Jansen}
\affiliation{
DESY, Zeuthen, Platanenallee 6, D-15738 Zeuthen, Germany
}
\email{karl.jansen@desy.de}

\author{C. McNeile}
\affiliation{
Department of Physics and Astronomy,
The Kelvin Building,
University of Glasgow, Glasgow, G12 8QQ, U.K.}
\email{c.mcneile@physics.gla.ac.uk}

\author{C. Michael}
\affiliation{
        Theoretical Physics Division, Dept of Mathematical Sciences,
 University of Liverpool, Liverpool L69 3BX, U.K.}
\email{cmi@liverpool.ac.uk}

\author{C. Urbach}
\affiliation{
Humboldt-Universit\"{a}t zu Berlin,  Institut f\"{u}r Physik
Mathematisch-Naturwissenschaftliche Fakult\"{a}t I,
Theorie der Elementarteilchen / Ph\"{a}nomenologie,
Newtonstr. 15, 12489 Berlin Germany}
\email{Carsten.Urbach@physik.hu-berlin.de}
 
\begin{abstract} 
 We report results for the masses of the flavour non-singlet light
$0^{++}$, $1^{--}$, and $1^{+-}$ mesons from unquenched lattice QCD at
two lattice spacings. The twisted mass formalism was used with two
flavours of sea quarks. For the $0^{++}$ and   $1^{+-}$ mesons we look
for the effect of decays on the mass dependence. For the light vector
mesons we study the chiral extrapolations of the mass. 
We report results for the leptonic and transverse decay constants 
of the $\rho$ meson. We test the mass dependence of the KRSF relations.
\end{abstract}

\maketitle

\section{Introduction} \label{se:intro}

Modern unquenched lattice QCD calculations include the dynamics  of
light sea quarks (with pion masses below 300 MeV) and  use multiple
lattice spacings and volumes~\cite{Jansen:2008vs}.  This has allowed
calculations of many basic quantities of long lived hadrons that decay
via the weak force to be computed  to high accuracy. Of particular note
is that unquenched lattice QCD calculations are now making contact with 
the results of chiral perturbation theory 
calculations~\cite{Leutwyler:2008ma,Necco:2007pr}, particularly for
light pseudoscalar mesons.

There has been much less work on studying resonances with the latest
generation of lattice QCD calculations. Some of the most interesting
questions in light quark hadron spectroscopy are looking for glueball
degrees of freedom in the experimental $f_0$ mesons and looking for
experimental evidence for the exotic $1^{-+}$ mesons. There are new
experiments, such as Gluex~\cite{Meyer:2006az}  and
PANDA~\cite{Lutz:2009ff} that will start around 2015, that aim to study
hadronic resonances. The new hadronic physics experiments will require
results from lattice QCD to guide their searches for new hadrons. The
lattice results for light resonances have recently been  reviewed 
by~\cite{Prelovsek:2008qu,McNeile:2007fu,McNeile:2007qf,Liu:2008ee,Lang:2007mq,Gattringer:2007da}.

In this paper we test basic lattice QCD techniques to study the $b_1$,
$a_0$, and $\rho$ mesons. The observation of the decay of the $\rho$
meson has been a long goal of the  lattice
community. The issue of dealing with the  decay of the $\rho$ meson has
stopped many calculations of weak decays such as  $B \rightarrow \rho
\nu e$~\cite{Ball:2004rg}. In the case  of determining $\mid V_{ub}
\mid$ from the  semi-leptonic decay $B \rightarrow \rho \nu e$, the 
simplest thing is to just ignore this decay and focus on  $B \rightarrow
\pi \nu e$. However there are some very important reactions such as $B
\rightarrow K^\star \gamma$ and  $B \rightarrow \rho \gamma$  that
have no simple equivalent form factors with a meson that is stable
under strong decay. The effect of the strong decays on these lattice
calculations is an unknown systematic error. It is also important to
understand the effect of strong decay on the $\rho$ meson for
calculations relevant to g-2~\cite{Aubin:2006xv,Renner:2009by}. 

It has been proposed 
(see~\cite{Prelovsek:2008qu,McNeile:2007fu,McNeile:2007qf,Liu:2008ee} 
for a review)
that the $a_0$(980) contains tetraquark or molecular
degrees of freedom. It is interesting to see whether quark-antiquark
operators actually couple to this state in lattice QCD calculations.
Understanding whether the $a_0(980)$ is a tetraquark is important for
classifying the $f_0$ and $a_0$ mesons into $\overline{q}q$ or
$\overline{q}\overline{q}qq$  multiplets~\cite{Liu:2008ee}.

%
%

First we define some notation. We call the  lightest flavour non-singlet
states  from the lattice calculations with $J^{PC}$ given by  $0^{++}$,
$1^{+-}$, and $1^{--}$  as the $a_0$, $b_1$, and $\rho$ mesons
respectively at the masses used in the lattice calculation. We include
the mass of the state when we deal with the experimental state, such as
$a_0(980)$, $\rho(770)$.

The plan of the paper is thus. We first discuss some general issues
about the effect of hadronic decays on mesons. We then describe the
details of the lattice QCD calculation and report results for the masses
in lattice units.  In section~\ref{se:Presults}  we discuss the
interpretation of the results for the  $a_0$ and $b_1$ channels. In
section~\ref{se:Sresults} we then discuss the results for the masses of
the vector mesons. We then discuss the leptonic decay constant of the
$\rho$  meson. 
In the penultimate section we test the KRSF relations.
In the final section~\ref{SE:conclusions} we draw our
conclusions.

\section{Generic background to the calculation} \label{se:background}

\begin{figure}
\centering
\includegraphics[%
  scale=0.3,
   angle=0,
  origin=c,clip]{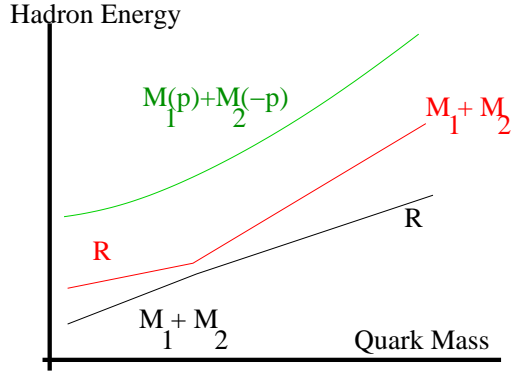}
\caption{The effect of decays on the energy levels of a resonance at 
finite volume.}
\label{fig:DECAYcartoon}
\end{figure}

At first analysis, it is not clear that the concept of a hadronic resonance makes
sense in an Euclidean lattice QCD calculation with a finite box size.
Naively, the size of the decay width could be a measure of the
systematic  error on the mass of the resonance on the lattice, however
there are arguments that suggest this is a pessimistic estimate.
 Michael~\cite{Michael:2006hf} reviews some of the phenomenology of
unstable  hadrons and notes that many unstable mesons  fit well with
mesons that are stable under the strong decays, using SU(3) symmetry for
example.
 Also Bijnens et al.~\cite{Bijnens:1997ni} obtained acceptable fits to
the masses of the light vector mesons with an effective theory (but some
parameters coming from a model) that didn't include the effect of the 
vector meson decay.


In figure~\ref{fig:DECAYcartoon} we show a ``picture'' of what  we
expect happens when a resonance ($R$) decays  to two mesons $M_1$ and
$M_2$ in the  lattice calculation. When the mass of the decay channels
and  the resonance are close there is mixing between them (an avoided
level crossing).
 The hadronic decay in figure~\ref{fig:DECAYcartoon} requires the
creation of a quark- anti-quark pair, so it is only present in
unquenched lattice QCD calculations.

For an S-wave decay the threshold for decay is  $M_R = M_{1}+M_{2}$. For
a P-wave decay the decay products must carry momentum. For example, in
the real world the $\rho$ decays into two pions, via a P-wave decay. The
threshold for decay at rest is $2 \sqrt{m_\pi^2 + (\frac{2 \pi}{L})^2 }$
where $L$ is the side of the box, assuming periodic boundary conditions
in space.  The CERN group~\cite{DelDebbio:2006cn,DelDebbio:2007pz} found
excited masses for the $\rho$ channel that were  consistent with $2
\sqrt{m_\pi^2 + (\frac{2 \pi}{L})^2 }$. For heavy quark masses it can be
more kinematically favourable  to study the decay of the $\rho$ meson
with one unit of momentum to decay to a pion at rest and a pion with one
unit of 
momentum~\cite{Rummukainen:1995vs,McNeile:2002fh,Aoki:2007rd}. It may well be that
one of the mesons ($M_1$ or $M_2$) in figure~\ref{fig:DECAYcartoon} is
also a resonance, in that case there will be second decay.  One example
of this is one of the decays of the $b_1$ meson.

 \begin{equation}
 b_1 \rightarrow \omega \pi \rightarrow \pi (\pi \pi \pi)
 \end{equation}

Our lattice calculations can in principle test the effect of the opening
of decay thresholds, because as we lower the sea quark masses in the
calculations, the various decays channels become open. In practice it
may be hard to see the effect of the  open decay as the quark mass
changes, because other systematic errors may change as well.

Although it appears that S-wave decays are kinematically 
easier to observe than P-wave decays, the $a_0$ and $b_1$
mesons are noisier than the $\rho$ meson.
 The $\rho$ meson at rest is stable to two pion decay in this
calculation, so for this state we try to build in the physics of the
meson decay by studying the chiral extrapolation formulae  in 
section~\ref{se:Sresults}. We also estimate the decay transition 
amplitude directly on the lattice, to gain an understanding of possible 
consequences of the mixing of the $\rho$ meson with the two pion state.

The MILC collaboration claimed to see some evidence
for the $a_0$ resonance to decay into two light 
hadrons~\cite{Bernard:2001av}.
Latter work showed that more analysis was required
to understand the $a_0$ decay in staggered 
calculations~\cite{Bernard:2007qf,Aubin:2004wf,Gregory:2005yr}.

L\"{u}scher has developed a technique to compute
the scattering phase shifts~\cite{Luscher:1991cf}. The method
was applied to 2-d theories~\cite{Gattringer:1992yz}
and the $\phi^4$ theory~\cite{Gockeler:1994rx}.
We have not investigated newer methods~\cite{Gockeler:2008kc,Durr:2008zz}
 based on 
L\"{u}scher's technique~\cite{Luscher:1991cf}, but plan to do
so in the near future.
Morningstar~\cite{Morningstar:2008mc} has recently presented a simple example 
of the basic method in quantum mechanics~\cite{DeWitt:1956be}.

\section{Details of the lattice calculation} \label{se:details}

Our lattice calculation uses the twisted  mass QCD
formalism~\cite{Frezzotti:2000nk}. Once a single parameter has been
tuned, twisted mass QCD has non-perturbative $O(a)$ 
improvement~\cite{Frezzotti:2003ni}. 
We call this maximally twisted mass QCD (MTMQCD).
This $O(a)$ improvement was checked
numerically by scaling studies using quenched QCD 
calculations~\cite{Jansen:2005gf,Jansen:2003ir,Petry:2008rt,AbdelRehim:2005gz,AbdelRehim:2006ve},
and has recently been checked in lattice perturbation 
theory~\cite{Cichy:2008gk}. As a prerequisite for large scale unquenched
calculations, the phase structure of twisted mass QCD has been 
studied~\cite{Farchioni:2004us,Farchioni:2005bh,Farchioni:2005tu,Farchioni:2004fs,Chiarappa:2006ae}.
The twisted mass formalism has recently been reviewed by
Shindler~\cite{Shindler:2007vp}. 

The ETM collaboration has already published a comparison of the lattice
results for $m_\pi$ and $f_\pi$ against  chiral perturbation
theory~\cite{Boucaud:2007uk,Boucaud:2008xu}. 
 Results for the nucleon and  $\Delta$ masses and a comparison with
chiral perturbation theory are reported in~\cite{Alexandrou:2008tn}. The
masses of the flavour singlet pseudoscalar mesons have been
presented~\cite{Michael:2007vn}.
  Light quark masses and decay constants from a partially quenched
analysis have been published from this data set~\cite{Blossier:2007vv}.
 There are ongoing projects to look at the  moments of parton
distributions~\cite{Capitani:2005jp,Baron:2007ti}, the form factor of
the pion~\cite{Frezzotti:2008dr}, and the properties of heavy-light
mesons~\cite{Blossier:2009bx}. For an overview of the broad range of
physics projects undertaken by the ETM collaboration see the review by
Urbach~\cite{Urbach:2007rt}.

For the gauge fields  we use the  tree-level Symanzik improved
gauge action~\cite{Weisz:1982zw}, which includes the
plaquette term $U^{1\times1}_{x,\mu,\nu}$ and rectangular $(1\times2)$ Wilson 
loops $U^{1\times2}_{x,\mu,\nu}$
 \begin{equation}
  \label{eq:Sg}
    S_g =  \frac{\beta}{3}\sum_x\Biggl(  b_0\sum_{\substack{
      \mu,\nu=1\\1\leq\mu<\nu}}^4\left \{1-\re\tr(U^{1\times1}_{x,\mu,\nu})\right \}\Bigr. 
     \Bigl.+
    b_1\sum_{\substack{\mu,\nu=1\\\mu\neq\nu}}^4\left \{1
    -\re\tr(U^{1\times2}_{x,\mu,\nu})\right \}\Biggr)\,  
 \end{equation}
 with  $b_1=-1/12$ and  $b_0=1-8b_1$. 
This choice of gauge action was made after a study of the 
phase structure of unquenched QCD with $n_f$=2 mesons.

The fermionic action for two degenerate flavours of quarks
 in twisted mass QCD is given by
 \be
S_F= a^4\sum_x  \bar{\chi}(x)\bigl(D_W[U] + m_0 
+ i \mu \gamma_5\tau^3  \bigr ) \chi(x)
\label{S_tm}
 \ee
 with   $\tau^3$ the Pauli matrix acting in
the isospin space, $\mu$ the bare twisted mass 
and the massless Wilson-Dirac operator given by 
 \be
D_W[U] = \frac{1}{2} \gamma_{\mu}(\nabla_{\mu} + \nabla_{\mu}^{*})
-\frac{ar}{2} \nabla_{\mu}
\nabla^*_{\mu} \quad.
 \ee
 where
 \be
\nabla_\mu \psi(x)= \frac{1}{a}\biggl[U^\dagger_\mu(x)\psi(x+a\hat{\mu})-\psi(x)\biggr]
\hspace*{0.5cm} {\rm and}\hspace*{0.5cm} 
\nabla^*_{\mu}\psi(x)=-\frac{1}{a}\biggl[U_{\mu}(x-a\hat{\mu})\psi(x-a\hat{\mu})-\psi(x)\biggr]
\quad .
 \ee
 Maximally twisted Wilson quarks are obtained by setting the untwisted
quark mass $m_0$ to its critical value $m_{\rm cr}$, while the twisted
quark mass parameter $\mu$ is kept non-vanishing in order to work away
from the chiral limit.
 In \eq{S_tm} the quark fields $\chi$ are in the so-called ``twisted
basis''. The ``physical basis'' is obtained for maximal twist by the simple
transformation
 \be
\psi(x)=\exp\left(\frac {i\pi} 4\gamma_5\tau^3\right) \chi(x),\qquad
\overline\psi(x)=\overline\chi(x) \exp\left(\frac {i\pi} 4\gamma_5\tau^3\right)
\quad.
\label{eq:maxTWIST}
 \ee
 In terms of the physical fields the action is given by
 \be
S_F^{\psi}= a^4\sum_x  \bar{\psi}(x)\left(\frac 12 \gamma_\mu 
[\nabla_\mu+\nabla^*_\mu]-i \gamma_5\tau^3 \left(- 
\frac{ar}{2} \;\nabla_\mu\nabla^*_\mu+ m_{\rm cr}\right ) 
+  \mu \right ) \psi(x)\quad.
\label{S_ph}
 \ee

 \begin{figure}
\begin{center}
\includegraphics[scale=0.8,angle=0]{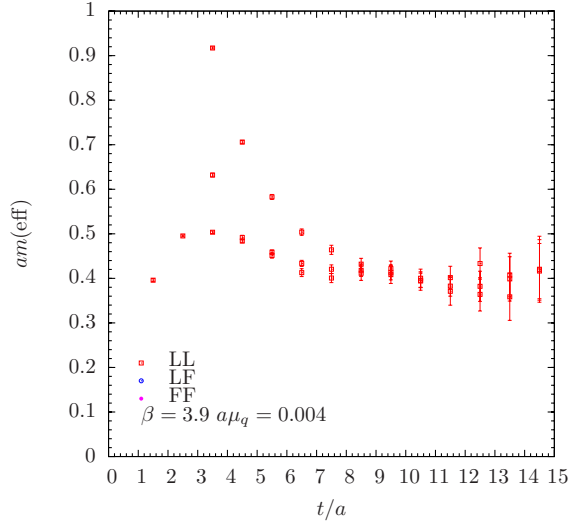}
\end{center}
 \caption {
Effective mass plot for the charged $\rho$ correlators (vector coupling)
for the $B_6$ ensemble. F and L are the fuzzed and local operators
respectively.
 }
\label{fig:rhomeff}
 \end{figure}

 The generation of the gauge configurations is  reported
in~\cite{Boucaud:2008xu,Urbach:2005ji,Jansen:2009xp}.
The methods used to extract the
masses and decay constants of the light mesons, from $n_f$=2 unquenched
twisted mass QCD are described in~\cite{Boucaud:2007uk,Boucaud:2008xu}.
Correlators separated by 10 trajectories were used.
The ensembles used in this calculation are summarized in
table~\ref{tb:ensembles}. We fit a matrix of correlators to a
factorising fit form~\cite{Boucaud:2008xu}. The basis of smearing
functions includes local and fuzzed operators. The correlators were
calculated with all-to-all quark propagators computed using the 
``one-end-trick''~\cite{McNeile:2006bz,Boucaud:2008xu}.

At finite lattice spacing, there is an order $a$ mixing of mesons  with
different parity in MTMQCD. When studying charged mesons this has the 
consequence that the $\rho$ and $a_1$ mesons mix. Assuming we are  at
maximal twist, the mixing will be of order $a$, then at large $t$ the
lightest state, the $\rho$ meson, will dominate. 
  The $\rho$ can be created by a  vector or tensor current  so we used a
4 by 4 matrix of correlators (vector/tensor and local/fuzzed). We obtain 
 a good fit with one meson state for $t/a > 7$. We checked these fits
using  either  a subset of operators or with more states.

The charged $a_0$ and $b_1$ mesons  mix under twisting with spin-exotic 
mesons so we do not expect at large $t$ any significant contributions
from parity  mixing since those states will be heavy.  For these cases, 
we fit   a 2 by 2  matrix of correlators (local/fuzzed) from $t/a > 3$
with two meson states.

In table~\ref{tb:rawData} we report the masses for the $a_0$, $b_1$ and
$\rho$ mesons in lattice units. In figure~\ref{fig:rhomeff} we plot the 
effective mass plot for the $\rho$ correlators for the  $B_6$ ensemble.

 In section~\ref{se:Presults} we process the raw data and convert the
results into physical units. To convert the results into lattice units
we use the scale from the pion decay constant, at
 $a_{\beta=3.9}$ = $0.0855(5)$ fm
 and 
 $a_{\beta=4.05}$ = $0.0667(5)$ fm. 
 These scales were consistent with those obtained from the  mass of the
nucleon~\cite{Alexandrou:2008tn}.

 \begin{table}[tb]
\caption{Summary of ensembles used in this calculation. The format of
the measurement column is number of blocks times block length.} 
\centering
\begin{tabular}{|c|c|c|c|c|} \hline
Ensemble & $\beta$ & $\mu$  &  $L^3 \times T$  &  Measurements \\ 
\hline
$B_1$ & 3.9  & 0.004  & $24^3  \times 48$  & $111 \times 8$ \\ 
$B_2$ & 3.9  & 0.0064 & $24^3  \times 48$  & $78 \times 32$ \\ 
$B_3$ & 3.9  & 0.0085 & $24^3  \times 48$  & $66 \times 32$ \\ 
$B_4$ & 3.9  & 0.01   & $24^3  \times 48$  & $38 \times 32$ \\ 
$B_5$ & 3.9  & 0.015  & $24^3  \times 48$  & $44 \times 32$\\ 
$B_6$ & 3.9  & 0.004  & $32^3 \times 64$ & $81 \times 6$ \\ 
$C_1$ & 4.05 & 0.003  & $32^3 \times 64$ & $64 \times 8$ \\ 
$C_2$ & 4.05 & 0.006  & $32^3 \times 64$ & $66 \times 8$ \\ 
$C_3$ & 4.05 & 0.008  & $32^3 \times 64$ & $61 \times 8$ \\ 
$C_4$ & 4.05 & 0.012  & $32^3 \times 64$ & $40 \times 8$ \\ 
\hline
\end{tabular}
\label{tb:ensembles}
 \end{table}

 \begin{table}[tb]
\caption{Masses in lattice units for the $a_0$, $b_1$, and $\rho$ mesons}
\centering
\begin{tabular}{|c|c|c|c|c|} \hline
Ensemble  & $a m_{b_1}$ & $a m_{a_0}$ & $a m_{\rho^+}$ & $a m_{\rho^0}$ \\
\hline
$B_1$ & 0.702(52) & 0.539(115) & 0.404(22) & 0.391(17) \\
$B_2$ & 0.685(28) & 0.573(59)  & 0.422(9)  & 0.434(17) \\
$B_3$ & 0.729(24) & 0.619(31)  & 0.428(8)  & 0.424(14) \\
$B_4$ & 0.681(29) & 0.666(34)  & 0.438(6)  & - \\
$B_5$ & 0.746(30) & 0.699(28)  & 0.481(7)  & - \\
$B_6$ & 0.674(29) & 0.636(53)  & 0.416(14) & 0.409(21) \\
$C_1$ & 0.552(38) & 0.509(45)  & 0.335(12) & 0.352(23) \\
$C_2$ & 0.555(29) & 0.410(29)  & 0.337(12) & 0.344(13) \\
$C_3$ & 0.526(35) & 0.511(26)  & 0.345(8)  & - \\
$C_4$ & 0.638(32) & 0.545(19)  & 0.368(6)  & - \\
\hline
\end{tabular}
\label{tb:rawData}
 \end{table}

In table~\ref{tb:rawData} we also include the lattice masses for the 
neutral $\rho^0$ operator. In the twisted mass formalism the 
 $\rho^0$ and  $\rho^+$ mesons are not degenerate because of the 
flavour violation from the twisted mass term. The results in
table~\ref{tb:rawData} show that  the  $\rho^0$ and $\rho^+$ are
essentially degenerate. A theoretical discussion with numerical examples
for why this is so, is  contained in~\cite{Frezzotti:2007qv}. 

As reported in~\cite{Frezzotti:2007qv} the main effect
of the flavour violation from the twisted mass term
is in the mass splitting between the mass of the $\pi^0$ and $\pi^+$ 
mesons. This has implications for decay thresholds of the 
$\rho^+$ and $\rho^0$ mesons.
Experimentally the dominant decays of the $\rho^+$ and
$\rho^0$ meson are to $\pi^+ \pi^0$ and $\pi^+ \pi^-$
respectively. The physical decay of $\rho^0$ to $\pi^0 \pi^0$
is not allowed, because of isospin symmetry, however
at non-zero lattice spacing this decay is allowed 
in twisted mass lattice QCD. At $\beta=$3.9 the mass splitting
between the $\pi^0$ and $\pi^+$ is approximately 
50 MeV at $\mu$=0.004~\cite{Boucaud:2008xu}. This should be compared with
one unit of quantised momentum of 600 MeV and 450 MeV on the
$24^3$ and $32^3$ lattices respectively at $\beta=3.9$.

\begin{figure}
\begin{center}
\includegraphics[scale=0.8,angle=0]{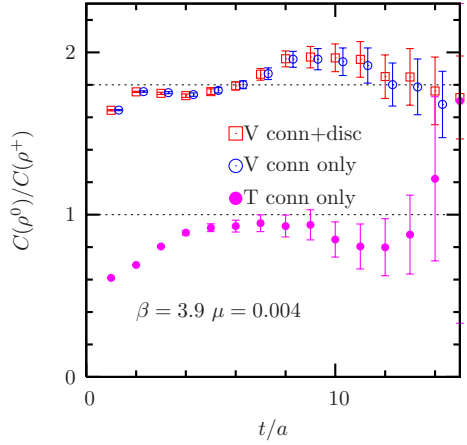}
\end{center}
\caption{
The ratio of correlators for the (i) connected 
neutral (ii) connected plus disconnected neutral to the
connected charged $\rho$ correlator for the 
$B_1$ ensemble. $T$ is the tensor and $V$ is the vector
current. As discussed in section~\ref{se:decay} the different
currents renormalise differently, which explains whether the ratio
tends to one or to the ratio of the square of the renormalisation factors.
}
\label{fig:B1neutraldis}
\end{figure}

\begin{figure}
\begin{center}
\includegraphics[scale=0.8,angle=0]{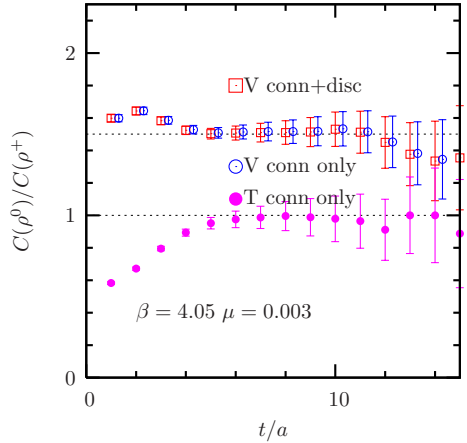}
\end{center}
\caption {
The ratio of correlators for the (i) connected 
neutral (ii) connected plus disconnected neutral to the
connected charged $\rho$ correlator for the 
$C_1$ ensemble. The notation is the same as for the caption of
figure~\ref{fig:B1neutraldis}.
}
\label{fig:C1neutraldis}
\end{figure}

 For a study of flavour singlet vector mesons such as the $\phi$ and
$\omega$, evaluation of  disconnected diagrams is required. Earlier
lattice work~\cite{McNeile:2001cr} showed that these contributions are
small.  For the  tensor coupling of the vector meson, the considerable
variance reduction possible using MTMQCD has allowed these contributions
to be evaluated  with some precision for the first
time~\cite{McNeile:2009mx} so yielding first principles 
results on the $\omega$-$\rho$ mass difference and mixing.

 Here we are discussing the flavour non-singlet mesons. 
 For the neutral $\rho$ meson there are also disconnected diagrams  that
contribute to the correlators, because the twisted mass formalism breaks
isospin symmetry at non-zero lattice spacing.
 These contributions would be expected to be small but, to check this, 
for the $B_1$ and $C_1$ ensembles we computed the  relevant disconnected
diagram for the vector mesons. Because of favourable variance
reduction~\cite{Boucaud:2008xu}, we are able to determine the 
disconnected contribution rather precisely for  neutral $\rho$
correlations using a vector coupling. 
The results  are in
figures~\ref{fig:B1neutraldis} and~\ref{fig:C1neutraldis}. 
As we explain in section~\ref{se:decay},
the neutral and charged vector currents renormalise differently,
thus explaining that the ratio of correlators tends to something
close to 2, rather than 1. The neutral and charged tensor current
renormalise the same way, so the ratio of correlators is close to
1.
For both
ensembles the  disconnected diagrams make a negligible contribution to
the correlators, so we do not consider their contribution any
further.

\section{Results for the masses of the $a_0$ and $b_1$
mesons}\label{se:Presults}

The results for the mass of the lightest flavour singlet  $0^{++}$ meson
from lattice QCD up to 2007 have been 
reviewed~\cite{Prelovsek:2008qu,McNeile:2007fu,McNeile:2007qf}. The
physics goal is to decide whether a $\overline{q}q$ interpolating
operator will couple to the experimental $a_0(980)$. The basic summary
of the older quenched work was that $\overline{q}q$ interpolating
operators did not see the  $a_0(980)$ meson and coupled to the higher
non-singlet state

 The unquenched calculation by the RBC 
collaboration~\cite{Prelovsek:2004jp} using $n_f$=2 domain wall fermions
also found a mass close to the mass of the experimental state
$a_0$(1450). In update on their analysis, that included 5 times the
statistics,  the RBC collaboration found 1.11(8) GeV for the lightest
state in the $0^{++}$ channel~\cite{Hashimoto:2008xg}.
 McNeile and Michael~\cite{McNeile:2006nv}, in an unquenched lattice QCD
calculation focused on the mass difference (in the hope that systematics
cancel), between the $1^{+-}$ and the $0^{++}$ mesons. Using this mass
splitting it was claimed that the lightest state in the $0^{++}$ channel
was consistent with the $a_0(980)$ state.
 Lang et al. reported masses for the lightest flavour non-singlet
$0^{++}$  consistent with the mass of the $a_0(980)$ meson, from an
unquenched lattice QCD calculation using chirally improved
fermions~\cite{Frigori:2007wa}.
 In an unquenched lattice QCD calculation with 2+1 flavours of sea quarks,
Lin et al.~\cite{Lin:2008pr} found that the lightest $a_0$
state to be consistent with the experimental $a_0(980)$.

\begin{figure}
\centering
\includegraphics[%
  scale=0.5,
 angle=270,
  origin=c,clip]{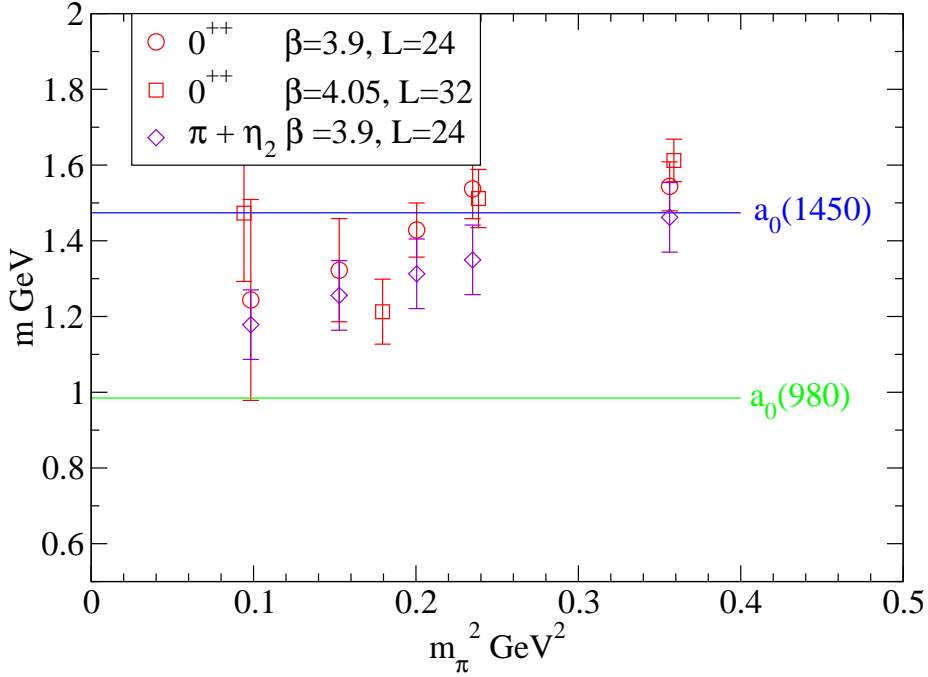}
\caption{Mass of lightest state in $0^{++}$ channel with the $\pi\eta_2$
decay threshold.}
\label{fig:ETMCa0}
\end{figure}

One complication is that experimentally the $a_0$ decays to $\pi \eta$.
In the two flavour world, the lightest $\eta$ meson is the flavour
singlet pseudoscalar meson  at the 800 MeV level. It is the mixing
between the light and strange loops in a lattice calculation that drives
the  mixing between flavour singlet pseudoscalar states $\eta$ and
$\eta'$. Hence, the decay thresholds will be very different for the 
$n_f$ = 2 and $n_f$ = 2+1 calculations that involve decay to a  flavour singlet
pseudoscalar meson. The ETM collaboration has recently published the
masses of the flavour singlet pseudoscalar meson (called $\eta_2$) on
these ensembles~\cite{Jansen:2008wv} and these results will be used to
estimate decay thresholds here. In figure~\ref{fig:ETMCa0} we plot the 
$a_0$ data and the decay thresholds.

To learn how to deal with mesons with open decays on the lattice, we
need some simple test cases to validate the lattice methods. 
 A bad example to study would be  the  $a_1(1260)$ because of its large 
experimental decay width of  250 to 600 MeV~\cite{Yao:2006px}.
 The $b_1(1235)$ meson  is good choice, because most models treat it  as
a $\overline{q}q$ state and its width is not too large at 142
MeV~\cite{Yao:2006px}. A direct study of the decay transition  $b_1 \to
\omega \pi $ has been made on the lattice with acceptable 
agreement~\cite{McNeile:2006bz} with the experimental decay width.
 To illustrate the impact of this (S-wave) decay threshold on the $b_1$ 
meson, we can use  the $\rho \pi$ decay threshold  (because the difference
between the $\rho$ and $\omega$ masses is shown to be
small~\cite{McNeile:2009mx}).

\begin{figure}
\centering
\includegraphics[%
  scale=0.5,
  angle=270,
  origin=c,clip]{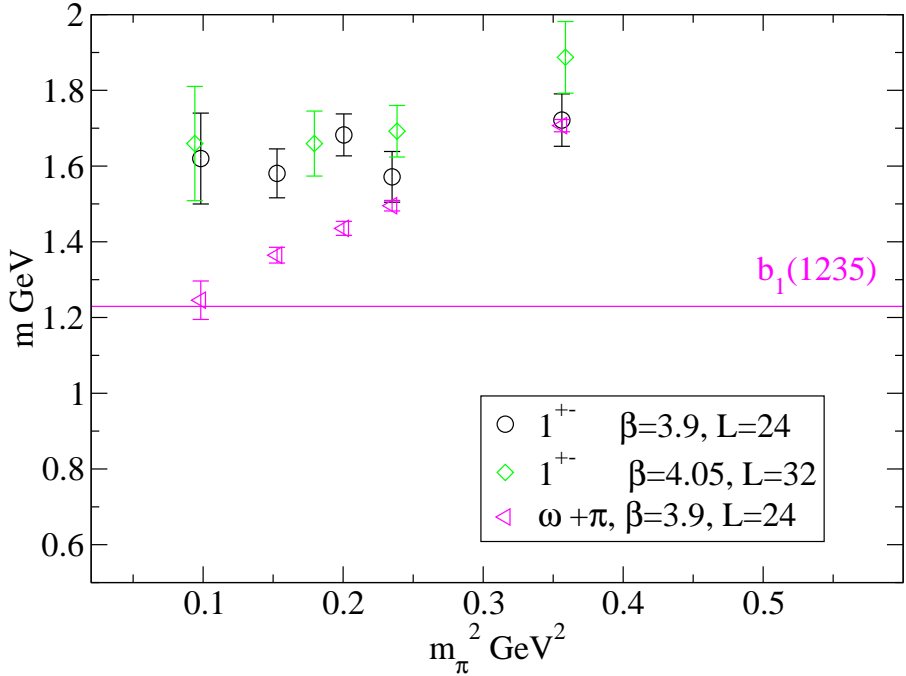}
\caption{Mass of $b_1$ state 
snd $\pi\omega$ threshold
as a function of square of pion mass.}
\label{fig:b1MASSETMC}
\end{figure}

In figure~\ref{fig:b1MASSETMC} we plot our results from the ETM
collaboration for the mass of the  $b_1$ meson with the estimate of the
$\omega \pi$ threshold, as a function of the square of the pion mass.
The mass of the lightest state in the $b_1$ channel is above
the decay threshold. 
This necessitates to include the
$\omega\pi$ operators
with the $b_1$ operators in a variational analysis,
which we plan to do in future work.

\begin{figure}
\centering
\includegraphics[%
  scale=0.5,
  angle=270,
  origin=c,clip]{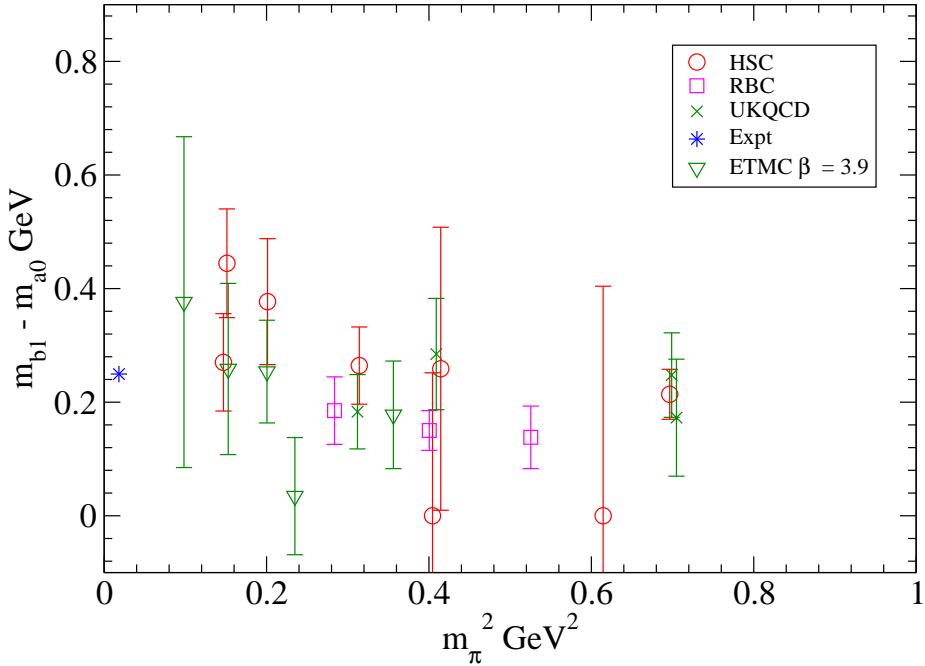}
\caption{Mass splitting between the $b_1$ and $a_0$ mesons. The plot
includes data from the RBC collaboration~\cite{Hashimoto:2008xg}, 
Hadron Spectrum Collaboration (HSC)~\cite{Lin:2008pr}, the 
UKQCD collaboration~\cite{McNeile:2006nv}, and the ETMC results from this work.}
\label{fig:b1a0split}
\end{figure}

In figure~\ref{fig:b1a0split} we plot the mass difference between the
mass of the $b_1$ and $a_0$ meson  as a function of the square of the
pion masses from a collection of recent unquenched lattice calculations.
 The fact that the majority of the results show the mass of the $a_0$
meson to be lighter than the  mass of the $b_1$ meson is good evidence for
the  lightest $a_0$ on the lattice corresponding to the experimental
$a_0(980)$ state. The Kentucky group have recently stressed that the
identification of $a_0(980)$ state on the lattice requires an
understanding of dynamics of the strong decay~\cite{Draper:2008tp}.


\section{Results for the masses of the light $1^{--}$ 
meson}\label{se:Sresults}

 In this section we will discuss the physical results for the mass of
the vector mesons. There is much more information on effective field
theory for the vector mesons, so there  is more we can do with the
chiral extrapolations in the mass of the  light quarks.
 The data for the $\rho$ meson are useful for applications such as the
calculation of the vacuum polarization tensor that is part of the QCD
corrections to $g-2$~\cite{Aubin:2006xv,Renner:2009by} and the
comparison of the electromagnetic form factor of the pion with the
vector exchange model~\cite{Frezzotti:2008dr}.

\begin{figure}
\centering
\includegraphics[%
  scale=0.5,
  angle=270,
  origin=c,clip]{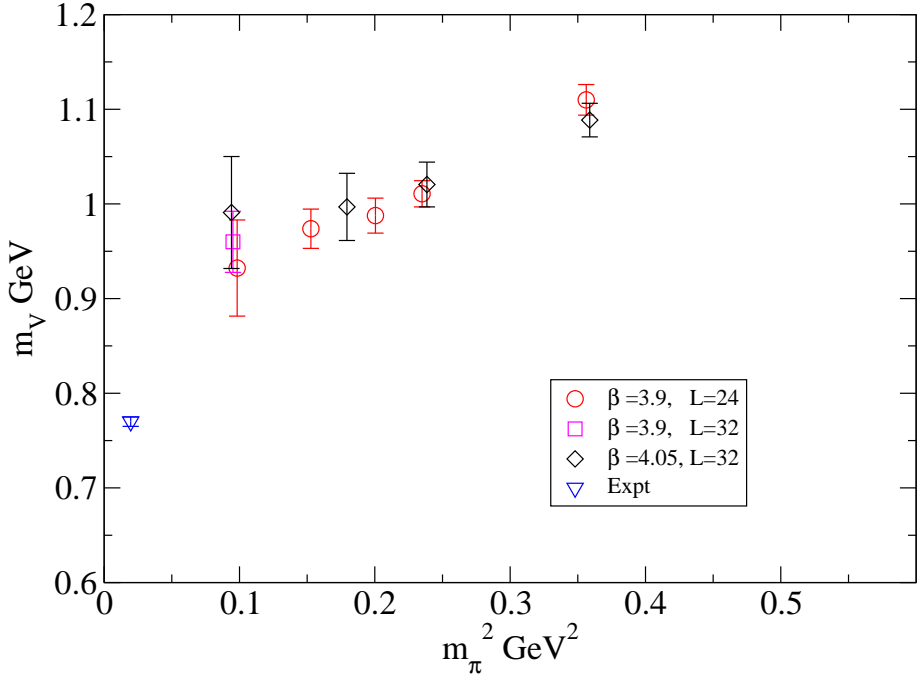}
\caption{The mass of the light vector meson as a function
of the square of the light pseudoscalar meson.}
\label{fig:rhoLatt}
\end{figure}

In figure~\ref{fig:rhoLatt} we plot the mass of the  lightest vector meson
as a function of the square of  the pion mass. Our lattice data seem
high relative to the  experimental mass of the $\rho$ meson. A more
detailed comparison with experiment requires a discussion of the chiral
extrapolations. Also the effect of $\rho$ decay needs to be accounted
for. 
 There has been a long history of attempts to
deal theoretically with the effect of the $\rho$ decay on the mass  of
the $\rho$ 
meson~\cite{DeGrand:1990ip,Leinweber:1993yw,Leinweber:2001ac,Allton:2005fb,Armour:2005mk}.

In~\cite{McNeile:2007fu} the vector meson mass as a function of the
square of the pion mass, was plotted with data from lattice QCD
calculations that used improved staggered (MILC
collaboration~\cite{Bernard:2001av}), and domain wall fermions
(RBC-UKQCD~\cite{Allton:2007hx}). There was reasonable agreement
between the data from the different formalisms, although the
statistical errors need to be reduced on some results (including ours).


Lattice correlators should have a  signal to noise
ratio which goes like $e^{-(m_M - m_\pi)t}$ for a meson of mass
$m_M$~\cite{DeGrand:2006zz}. We have checked that our data at $\beta$ =
3.9 obeys this relation. So there is no fundamental problem with the
increase in the statistical errors as the mass of the light quarks is
reduced. On a subset of the  configurations we tried a technique called
color dilution to improve the signal to noise ratio for the connected
$\rho$ correlators~\cite{Foley:2005ac}. This did not reduce the
statistical noise. ETMC have used an extrapolation of the partially
quenched $\rho$ masses to reduce the statistical
errors~\cite{Dimopoulos:2008hb}.

\begin{figure}
\centering
\includegraphics[%
  scale=0.5,
  angle=270,
  origin=c,clip]{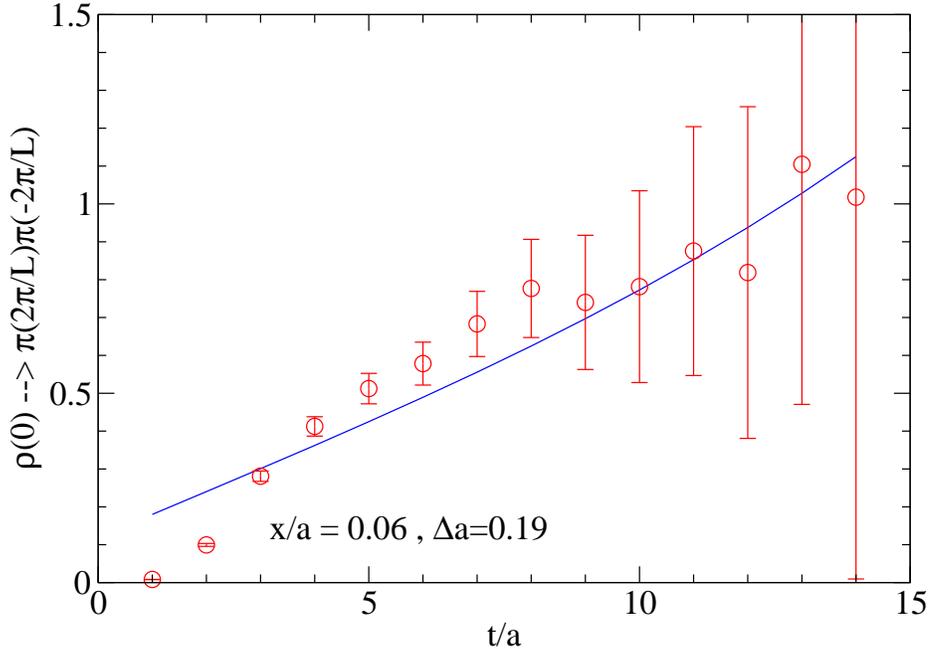}
\caption{The correlator $c_3(t)$ in equation~\ref{eq.xt} as a function of time.}
\label{fig:rhoDECAY}
\end{figure}

At $\beta = 3.9$ and $\mu = 0.004$, we have also estimated the mixing 
element between $\rho^0$ and $\pi^+ \pi^-$ from a correlator ratio using
the method described in~\cite{McNeile:2002fh}. 
 The three point function ratio  was computed using
 \begin{equation}
c_3(t) = 
  { \langle \rho(0) | \pi(t) \pi(t) \rangle  \over
 \langle \rho(0) | \rho(t) \rangle^{1/2} 
\langle \pi(0) \pi(0) | \pi(t) \pi(t) \rangle^{1/2}  }
 \label{eq.xt}
 \end{equation}
 When the $\rho$-mass and $\pi \pi$ energy are degenerate, for small
enough $x$~\cite{McNeile:2002fh} (where $x=\langle \rho \mid \pi\pi\rangle$), 
this ratio can be fitted to  the model in eq.~\ref{eq.xtmodel}.
 \begin{equation}
c_3(t) \rightarrow 
 xt + {\rm const}
 \label{eq.xtmodel}
 \end{equation}
The formalism required, where the $\rho$-mass and $\pi \pi$ energy are not degenerate,
is discussed in~\cite{McNeile:2000xx}.
 The correlator ratio $c_3(t)$ is plotted in figure~\ref{fig:rhoDECAY}
for the decay $\rho \rightarrow \pi(k=\frac{2 \pi}{L})\pi(k=-\frac{2
\pi}{L})$. Since the $\rho$-mass is somewhat larger (by 0.19 in lattice
units)  than the lightest two pion energy, we plot in the figure a
theoretical curve which  modifies eq.~\ref{eq.xtmodel}, taking this into
account, as used in ref.~\cite{McNeile:2002fh}. 
 This fit to the three point function ratio gives $ax = 0.060(15) $.
Since on a lattice, energy is not conserved,  we  have evaluated the
transition amplitude to a final state with sufficient momentum that
its energy is  more than that of the $\rho$ at rest, so strictly   a
zero decay width. So, to compare with experiment, it is optimum to 
evaluate the coupling constant. This may have some dependence on
momentum  in general, but it is a useful point of reference. 
 The $g^2_{\rho \pi \pi}$ coupling defined via
\begin{equation}
\Gamma = \frac{g^2_{\rho \pi \pi} }{6 \pi}  \frac{k^3 }{m_{\rho}^2}
\end{equation}
 is found to be $g_{\rho \pi \pi}$=5.2(1.3). The corresponding value of 
$g_{\rho \pi \pi}$ from the experimental value of the $\rho$ width is
6.0. So we have consistency between the lattice estimate of the 
coupling between $\rho$ and $\pi \pi$ and that observed.

Since we measure the strength of the transition from $\rho$ to $\pi \pi$ on 
the lattice (namely $x$), we can estimate the mass shift caused by this mixing.
Then with a two-state model with energy difference $\Delta$ where 
 \begin{equation}
 \Delta = E_2 - E_1 = 2 \sqrt{m_\pi^2 + (\frac{2 \pi}{L})^2} - m_\rho
\end{equation}
 with $a \Delta=0.19$ in our case,
 we obtain, using~\cite{McNeile:2002fh}, a shift (downwards for the 
 $\rho$) of 
 \begin{equation}
m_{split} = \sqrt{ \Delta^2/4 + x^2} - \Delta/2
 \end{equation}
 This mixing produces a 4\% downward shift in the mass of the  $\rho$
for ensemble $B_1$ using this simplified mixing scheme.
This shift is comparable to our statistical error for that state.
 This suggests that the  mass of the vector mesons in
figure~\ref{fig:rhoLatt} are largely unaffected by the two $\pi$ decay.

 This mixing argument can be used to compare expectations between  the
$B_6$ ensemble with $L=32$ and that above with $L=24$ above. The
differences will be that  the energy gap  will become much smaller
($\Delta=0.08$) since the minimum momentum  is reduced while the mixing
contribution ($x^2$) will be reduced proportionally to the  spatial
volume. The net effect is a rather similar estimate  which is consistent
with our results which show that the $\rho$ mass  from the $B_6$
ensemble is  half-$\sigma$  higher than for the $B_1$ ensemble.



We now discuss the chiral extrapolation of the vector masses to the
physical point. 
 For the case of an effective field theory for vector mesons, the issues
in writing down an effective field theory are less clear than for pions.
 A fully relativistic Lagrangian can be used for the vector fields or a
heavy meson effective theory
(HMET)~\cite{Jenkins:1995vb,Bijnens:1997ni}.  The connection between the
different effective theories is discussed 
in~\cite{Meissner:1987ge,Bijnens:1997ni}.

The most basic effective field theory  for the light vector meson
predicts that the  mass of the vector meson depends on the mass of the 
pion via~\cite{Jenkins:1995vb,Bijnens:1997ni}:
 \begin{equation}
M_\rho = M^0_\rho + c_1 M_\pi^2  + c_2 M_\pi^3 
\label{eq:MVsimple}
 \end{equation}

The pions involved in $\rho$ decay are not soft so $\rho \rightarrow
\pi \pi$ can not be studied using chiral perturbation theory 
with power counting~\cite{Jenkins:1995vb,Bijnens:1997ni}.
 However, Bijnens et al.~\cite{Bijnens:1997ni} successfully  fitted the
masses of the light vector mesons $\rho$ to $\phi$, including
electromagnetic effects, using HMET but not including the dynamics of
the $\rho \rightarrow \pi \pi$ decay.

The $\rho$ decay will effect the chiral extrapolation model
used to extrapolate the mass of the $\rho$ meson. The Adelaide
group have studied different 
regulators~\cite{Leinweber:2001ac,Allton:2005fb,Armour:2005mk} 
for the effective field theory of $\rho$ decay. This produced 
additional mass dependence at very light pion masses.

Models for the effect of $\pi \omega$ and $\pi \pi$ contributions to 
the mass of the $\rho$ meson have direct implications for the mass 
of the $\omega$ meson (which has Â$\pi \rho$ contributions). 
 Hence lattice results for the quark dependence of the mass splitting of
the  $\omega$ to $\rho$  mesons~\cite{McNeile:2009mx} allow
further constraints to the study of individual terms.

 Bruns and Mei{\ss}ner~\cite{Bruns:2004tj} have published a chiral
extrapolation formulae for the mass of the $\rho$ meson. The derivation
used a modified $\overline{MS}$ regulator and a power counting scheme.
 \begin{equation}
M_\rho = M^0_\rho + c_1 M_\pi^2  + c_2 M_\pi^3 + 
c_3 M_\pi^4 \ln ( \frac{M_\pi^2}{M_\rho^2} )
\label{eq:brunsmeis}
 \end{equation}
 The term with the $c_3$ coefficient is due to the  self energy (in the
infinite volume limit).
 Bruns and Mei{\ss}ner~\cite{Bruns:2004tj} recommend that  the size of
the $c_i$ coefficients obtained from the fits to the lattice
calculations be checked against constraints from low energy effective
constants.
 However they only
quote, as reasonable, the constraints that $\mid c_i \mid < 3$.
The Adelaide group~\cite{Leinweber:2001ac} claimed to know
the sign and magnitude of the $c_2$  coefficient ($c_2 \sim$ $-1.70$
$GeV^{-2}$),  but 
 Bruns and Mei{\ss}ner~\cite{Bruns:2004tj} claim their bounds are
more general.

Using one loop chiral perturbation theory and a technique called the 
inverse amplitude method, Hanhart 
et al.~\cite{Hanhart:2008mx,Rios:2009az} estimate
$c_1$ = $0.90 \pm 0.11 \pm 0.13$ $\mbox{GeV}^{-1}$
$M^0_\rho = 0.735 \pm 0.0017$ GeV.

Bruns and Mei{\ss}ner~\cite{Bruns:2004tj} from an analysis of an old
lattice QCD calculation by the CP-PACS 
collaboration~\cite{Aoki:1999ff}, found that the curvature from the 
non-analytic terms can produce either an increase or decrease in the
vector mass over a simple linear fit. CP-PACS used the string tension
($440 $ MeV) to set the lattice  spacing~\cite{Aoki:1999ff}, this
corresponds to $r_0 \sim$ 0.54 fm, roughly 10 \% higher than the
preferred $r_0$ from the pion decay constant. If there is any ambiguity
in the  lattice spacing, then this can hide the curvature from the 
non-analytic terms.

Unfortunately the size of errors on the $\rho$ data and the number
of points does not allow us to include the $c_2$ and $c_3$ coefficients as
free parameters. To get some idea of the effect of these terms  we
use the augmented $\chi^2$ 
method~\cite{Lepage:2001ym,Morningstar:2001je} where the physics
constraints from Bruns and Mei{\ss}ner~\cite{Bruns:2004tj} can be built 
into the fit with Bayesian techniques. The augmented $\chi^2$ is used to
constrain $c_2$ and  $c_3$.
 \begin{equation}
\chi^2_{aug} = \chi^2 + \sum_{j=2}^3 \frac{ (c_i -0)^2 } {3^2} 
\label{eq:augchisq}
 \end{equation}
 Schindler and Phillips have recently discussed using an augmented
$\chi^2$ to using information from effective theories in chiral
extrapolations of lattice data. We use the bootstrap method to estimate
the errors. In principle given the probability distribution, the errors on
the parameters can be obtained by integrating  the Monte Carlo
integrals~\cite{Morningstar:2001je,Schindler:2008fh}. Chen et al.
checked~\cite{Chen:2004gp} that consistent errors were obtained from a
bootstrap analysis and from an error analysis based on the  augmented
$\chi^2$ being a quadratic function of the fit parameters around the
minimum.

We also investigated an approach developed by the 
Adelaide~\cite{Leinweber:2001ac} group. 
The Adelaide method uses a dipole regulator, rather than
the $\overline{MS}$ scheme, to regulate the effective field
theory corrections to the $\rho$ mass~\cite{Leinweber:2001ac}. 
The extrapolation
model for the mass of the $\rho$ meson is
 \begin{equation}
M_\rho = M_\rho^0 + c_1 M_\pi^2 + 
\frac{
\Sigma_{\pi \omega}(\Lambda_{\pi \omega},M_\pi)
+
\Sigma_{\pi \pi}(\Lambda_{\pi \pi},M_\pi)    
}
{2( M_\rho^0  + c_1 M_\pi^2  ) }
\label{ad:rhoFIT}
 \end{equation}
where $\Sigma_{\pi \omega}$ and $\Sigma_{\pi \pi}$
are the self energies from the $\pi \pi$ and $\pi \omega$ states.
The fit parameters in equation~\ref{ad:rhoFIT} are
$M_{\rho}^0$, $c_1$ and $\Lambda_{\pi \omega}$. The parameter
$\Lambda_{\pi \pi}$ is related to  $\Lambda_{\pi \omega}$.
The $\rho$ self energy contribution $\Sigma_{\pi \pi}$
contains a cut at $m_\rho = 2 m_{\pi}$ for 
the decay  $\rho \rightarrow \pi \pi$. For the continuum
integral we used the principle value of the integral
when the decay is open. We found that our data was too noisy
to get stable fits from this method. We were also unable to resolve
the quadratic $c_2$ term in equation~\ref{eq:MVsimple}, because the
error bars were too large.
The original study~\cite{Leinweber:2001ac} 
of equation~\ref{ad:rhoFIT} used
$\rho$ masses from lattice QCD with 1\% errors at a heavier
quark masses~\cite{Leinweber:2001ac}.

The summary of the final results is in  table~\ref{tb:finalFITS}. We use
the pion mass of 135 MeV, because we don't include any electromagnetism
in  the lattice calculation. We also extrapolate our results to mass of
the notional strange-strange pseudoscalar meson (696 MeV). We call  this the
unitary $\phi$ analysis. Note that a better approach to the $\phi$ meson 
within an $n_f=2$ formalism would be to treat the strange quark as a 
(partially quenched) valence quark with a sea of light quarks.

\begin{figure}
\centering
\includegraphics[%
  scale=0.5,
  angle=270,
  origin=c,clip]{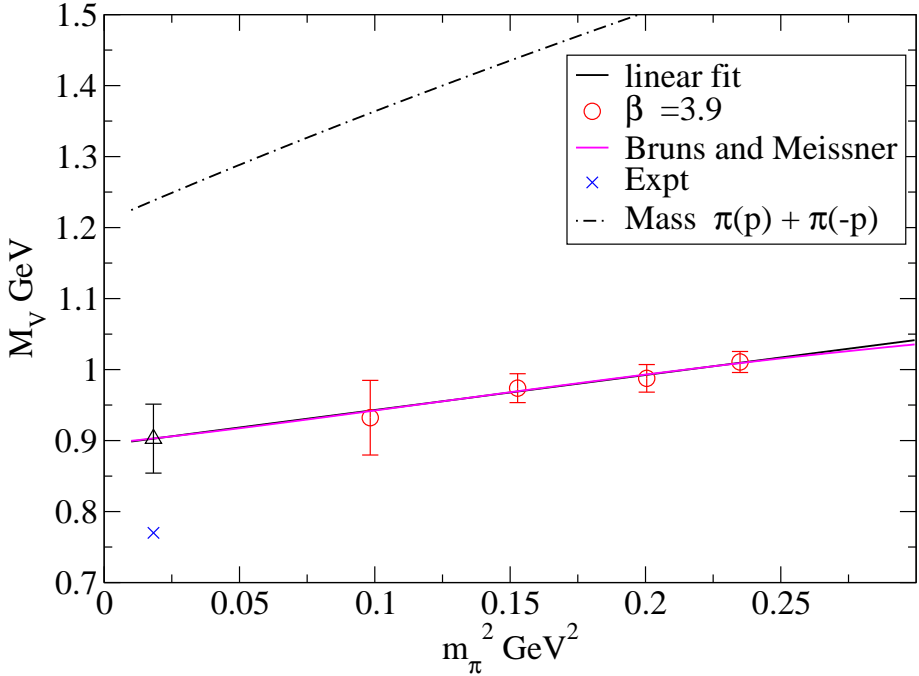}
\caption{Fit to the mass of the vector 
meson using a linear fit in the square of the pion mass
and equation~\ref{eq:brunsmeis} at $\beta=3.9$. Also included in the plot is
the first decay threshold to $\pi\pi$ for L=24.}
\label{fig:adelaie}
 \end{figure}

In figure~\ref{fig:adelaie} we plot the linear fit and the extrapolation
model in equation~\ref{eq:brunsmeis}.
 \begin{table}[tb]
\caption{The $\rho$ mass from chiral extrapolation from different
fit models at $\beta$ = 3.9 }
\centering
 \begin{tabular}{|c|c|c||c||c|c|c|c|} \hline
 Equation & Model   & $m_\rho$  GeV  & $m_\phi$  GeV  &  $M^0_\rho$
 GeV  
&  $c_1$ ${GeV}^{-1}$ 
& $c_2$ $(GeV)^{-2}$& $c_3$ $(GeV)^{-3}$   \\ \hline
\ref{eq:MVsimple} & linear & 0.90(4) & 1.13(8)   & 0.89(5) & 0.49(26) & -
& -  \\  
\ref{eq:brunsmeis} & Bruns and Mei{\ss}ner &
0.90(5) & 1.07(10)  & 0.89(6) & 3.5(4.7) & -0.09(81) &
-0.82(41) \\  
\hline
\end{tabular}
\label{tb:finalFITS}
 \end{table}

The lattice data for the vector mesons seem  to prefer a smaller lattice
spacing than the scales obtained from the pion decay
constant~\cite{Boucaud:2007uk}  and the nucleon
mass~\cite{Alexandrou:2008tn}, this is probably because we are missing
some of the effect from the $\rho$ decay and possibly also from the
dynamical strange  quark.

\section{The decay constants of the $\rho$ and $\phi$ mesons} \label{se:decay}

We first introduce the leptonic  decay constant of the vector mesons,
such as the  $\rho$ or $\phi$, in the continuum~\cite{Lewis:1996qv}. The
decay constant of the vector meson $V$ is defined~\cite{AliKhan:2001tx}
via
 \begin{equation}
\langle 0 \mid V_\mu \mid V \rangle = 
m_\rho f_V
\epsilon_\mu
\label{eq:rhodecayDEFN}
 \end{equation} 
 where the vector current is defined via
 \begin{equation}
V_{\mu}(x) = \overline{\psi}(x) \gamma_\mu \psi (x)
 \end{equation} 
 There are other possible (slightly different)  definitions of the decay
constant of the $\rho$ definitions, for example as used by Lewis and
Woloshyn~\cite{Lewis:1996qv}.

The decay constants of the $\rho$ and $\phi$ mesons
can be extracted from $\tau$ decay and $e^+$ $e^-$
annihilation (see~\cite{Becirevic:2003pn,Bijnens:1996kg} 
for a discussion). 
 \begin{eqnarray}
f_{\rho^+}^{expt} & \sim & 208 \; \mbox{MeV} \\
f_{\rho^0}^{expt} & \sim & 216(5) \; \mbox{MeV} \\
f_{\phi}^{expt}   & \sim & 233 \; \mbox{MeV}
 \end{eqnarray}
 The difference between the experimental values of  the
$f_{\rho^+}^{expt}$ and $f_{\rho^0}^{expt}$ is probably due to the
problems of extracting the  parameters of the $\rho$ meson  from
experimental data, rather than  electromagnetic effects that are
important for light pseudoscalar mesons~\cite{Cirigliano:2007ga}.

The transverse decay constant ($f_V^T(\mu)$) of the $V$ meson is defined
by
 \begin{equation}
\langle 0 \mid 
\overline{\psi} \sigma_{\mu \nu} \psi
\mid V \rangle = 
i f_V^T(\mu) ( p_\mu \epsilon_\nu - p_\nu \epsilon_\mu )
\label{eq:TRANSrhodecayDEFN}
 \end{equation} 
 where $\sigma_{\mu \nu} = i/2 [\gamma_\mu , \gamma_\nu] $. It is
convenient to introduce the tensor  current $T_{\nu \mu} =
\overline{\psi} \sigma_{\mu \nu} \psi$. In the lattice calculations we
do not include any momentum. 

There is no experimental result for the tensor decay constant
$f_V^T(\mu)$ for the $\rho$ or $\phi$ mesons. However, light cone sum
rules require the transverse decay constant of the $\rho$ 
meson~\cite{Ball:2006nr,Ball:2006eu} for the extraction of $\frac{\mid
V_{td} \mid}{\mid V_{ts} \mid } $ from the $B \rightarrow \rho \gamma $
and  $B \rightarrow K^\star \gamma $ decays. The transverse decay
constant of the $\rho$ meson is also used in the analysis of other B 
decays~\cite{Ball:2004rg}. There have been previous lattice QCD 
calculations of the  transverse decay constants of  the $\rho$ 
meson~\cite{Braun:2003jg,Becirevic:2003pn,Gockeler:2005mh,Allton:2008pn}.

There needs to be a way to estimate the effect of the strong decay of
the $\rho$ meson to two $\pi$ on the decay constants, in the same way we
tried for the $\rho$ mass in section~\ref{se:Sresults}. A simple test is
look at the $f_V$ decay constant for the $\rho$ and $\phi$ mesons as
these give us an estimate of our accuracy. The majority of older lattice
QCD calculations concentrated on the ratio of $f_V^T$ to  $ f_V$.

There are various correlators that can be used to extract the $f_V$
and $f_V^T$ decay constants.  For example the correlators in
equations~\ref{eq:VV}, \ref{eq:VT} and~\ref{eq:TT}. Our results are
based on factorising fits to a basis of 4 by 4 smearing functions that
include the local operators as matrix elements in the 
smearing matrix, so the operators in
equations~\ref{eq:VV}, \ref{eq:VT} and~\ref{eq:TT} are included.

\begin{equation}
\sum_x \sum_{\mu=1}^{3}
\langle V_\mu(x,t_x) V_\mu(0,0)^\dagger \rangle
\rightarrow
\frac{3 m_V f_V^2 e^{-m_V t_x}  }{2 }
\label{eq:VV}
\end{equation}


\begin{equation}
\sum_x \sum_{\mu=1}^{3}
\langle T_{\mu 0} (x,t_x) V_\mu(0,0)^\dagger \rangle
\rightarrow
\frac{3  f_V f_V^{T} m_V e^{-m_V t_x} }{2}
\label{eq:VT}
\end{equation}

\begin{equation}
\sum_x \sum_{\mu=1}^{3}
\langle T_{\mu 0} (x,t_x) T_{\mu 0}  (0,0)^\dagger \rangle
\rightarrow
\frac{3 m_V (f_V^{T})^2 e^{-m_V t_x}  }{2 }
\label{eq:TT}
\end{equation}

The local vector $V_\mu$ and $T_{\mu \nu}$ tensor currents need to be
renormalised. This involves some discussion of the twisted mass
formalism. We do all our fits in the twisted bases, however  the
identification of states is done in the physical 
basis~\cite{Boucaud:2008xu}.
 Assuming that the calculations are done at maximal twist 
(see equation~\ref{eq:maxTWIST}), we have

\begin{eqnarray}
\langle i \mid V_\mu^3 \mid j \rangle_{cont}
& = & Z_V \;
\langle i \mid V_\mu^{3} \mid j \rangle_{twisted \; lattice}
\\
\langle i \mid V_\mu^\alpha \mid j \rangle_{cont}
& = & Z_A \; \epsilon^{3\alpha\beta} \; 
\langle i \mid A_\mu^{\beta} \mid j \rangle_{twisted\; lattice}
\\
\langle i \mid T_{\nu\mu}^\alpha \mid j \rangle_{cont}
& = & Z_T \;
\langle i \mid T_{\nu\mu}^{\alpha} \mid j \rangle_{twisted \; lattice}
 \end{eqnarray}
 where $\alpha$ takes the values of 1 or 2. Given that we found that the
disconnected graphs for vector mesons were negligible (in
section~\ref{se:Presults}),  then the connected charged and neutral
vector mesons give us a separate estimate of the decay constants that
use different  renormalisation constants. This is a useful test of  the
renormalisation and cut off effects.

 \begin{table}[tb]
  \caption{
 Summary of the non-perturbative renormalisation 
factors used in this calculation. The $C(\mu)$ function is the solution,
in equation~\ref{eq:Cevolve},
of the RG equation for the tensor current}.
\centering
\begin{tabular}{|c|c|c|c|c|} \hline
$\beta$ & $Z_A$  &  $Z_T(\mu=\frac{1}{a})$ & $Z_V$ &  
$\frac{C(\mbox{2 GeV})}{C(\mu=\frac{1}{a})}$ \\
\hline
3.9     & 0.771(4) & 0.769(4) & 0.6104(02) & 1.01 \\
4.05    & 0.785(6) & 0.787(7) & 0.6451(02) & 1.03 \\
\hline
\end{tabular}
\label{tb:Zfactsummary}
 \end{table}

The relevant renormalisation factors $Z_V$, $Z_T$, and $Z_A$, have been
computed~\cite{Dimopoulos:2007fn,Dimopoulos:2009prep} using the Rome-Southampton
non-perturbative method~\cite{Martinelli:1994ty}. The  $Z_V$ factor has
also been computed using the conserved vector
current~\cite{Boucaud:2008xu}. It was found that the conserved vector
current produced a more accurate estimate of $Z_V$ than the
Rome-Southampton method, so we use the result from the conserved current
in this analysis. 
In this paper we use the $Z_A$ and $Z_T$ values
calculated through the 'p2-window' method without the use of the subtraction
of $O(a^2g^2)$ terms.
In table~\ref{tb:Zfactsummary} we summarise the
renormalisation factors used in this
calculation~\cite{Dimopoulos:2007fn,Dimopoulos:2009prep}. 

The value of the tensor current depends on the  scale. The tensor
current at $\mu_a^2$ is obtained from  that at another scale
($\mu_b^2$) by using the renormalisation group equation.
 \begin{equation}
Z_T( \mu_a^2) = \frac{C(\mu_a^2) }{C(\mu_b^2)}  Z_T( \mu_b^2)
 \end{equation}
 \begin{eqnarray}
C(\mu^2) & = & 
\left( \frac{\alpha_s(\mu)}{\pi} \right)^{\gamma_0}
[ 1 + 
\left( \frac{\alpha_s(\mu)}{\pi} \right)
(
\overline{\gamma}_1 - \overline{\beta}_1 \overline{\gamma}_0 
)
\nonumber \\
& + \frac{1}{2} & \left( \frac{\alpha_s(\mu)}{\pi} \right)^2
[
(
\overline{\gamma}_1 - \overline{\beta}_1 \overline{\gamma}_0 
)^2
+
\overline{\gamma}_2 + \overline{\beta}_1^2 \overline{\gamma}_0 
- \overline{\beta}_1 \overline{\gamma}_1
- \overline{\beta}_2 \overline{\gamma}_0
]
]
\label{eq:Cevolve}
 \end{eqnarray}
 with 
 \begin{equation}
\overline{\gamma}_i = \frac{\gamma_i }{\beta_0},
\;\;\;\;\;\;
\overline{\beta}_i = \frac{\beta_i }{\beta_0}
 \end{equation}
 \begin{eqnarray}
\beta_0 & = & \frac{1}{4} \left(11 - \frac{2}{3} n_f\right)  \nonumber \\
\beta_1 & = & \frac{1}{16} \left(102  -\frac{38}{3} n_f \right)  \nonumber \\
\beta_2 & = & \frac{1}{64}  \left(\frac{2857}{2} - \frac{5033 n_f}{18}  
+ \frac{325 n_f^2}{54} \right)
 \end{eqnarray}

The anomalous dimension for the tensor current has been computed by
Gracey~\cite{Gracey:2000am,Gracey:2003yr} to three loops in the 
$RI^\prime$ and the $\overline{MS}$ schemes.
\begin{eqnarray}
\gamma_0 & = &  \frac{1}{3}  \nonumber \\
\gamma_1 & = &   \frac{543 - 26 n_f  }{216}    \nonumber \\
\gamma_2 & = &  
-\left( \frac{36 n_f^2 + 1440 \zeta_3 n_f + 5240 n_f
+2784 \zeta(3) - 52555} {5184} \right)
\end{eqnarray}
 where the value of the standard constant is $\zeta_3$ = 1.20206.

For the coupling we used RunDec package~\cite{Chetyrkin:2000yt} to
compute the coupling from $\Lambda_{QCD}$  using 4-loop
evolution~\cite{vanRitbergen:1997va,Czakon:2004bu}. 
There has not been a calculation
of the strong coupling using information from these configurations. We
used the value of  $\Lambda_{QCD}$ = 261(17)(26) MeV from
QCDSF~\cite{Gockeler:2005rv}. The QCDSF value is consistent with that
from ALPHA~\cite{DellaMorte:2004bc}, that also used $n_f$ = 2 QCD.

 \begin{table}[tb]
  \caption{
Summary of the leptonic decay constant of the vector meson
for the different ensembles from this calculation.
}
\begin{center}
\begin{tabular}{|c|c|c|c|c|} \hline
 & \multicolumn{2}{|c|}{Charged} & \multicolumn{2}{|c|}{Neutral}\\ 
Ensemble &  $af_V / Z_A$ & $f_V$ MeV &  $f_V$ MeV &  $af_V / Z_V$ \\ \hline
$B_1$    & 0.13(1)   &  234(18)  & 252(13) & 0.179(9)  \\ 
$B_2$    & 0.148(4)  &  264(7)  & 283(18) & 0.20(1) \\ 
$B_3$    & 0.149(4)  &  265(7)  & 274(14) & 0.19(1)  \\ 
$B_4$    & 0.151(4)  &  269(7)  &  - &  -  \\ 
$B_5$    & 0.162(4)  &  289(11)  & - & - \\ 
$B_6$    & 0.151(7)   &  269(12)  & 275(19) & 0.19(1) \\  
$C_1$    & 0.119(10)  &  277(24)  & 306(22) & 0.16(1) \\  
$C_2$    & 0.117(6)   &  272(14)  & 291(14) & 0.152(7) \\ 
$C_3$    & 0.117(4)   &  272(10)  & - & -  \\  
$C_4$    & 0.121(3)   &  281(9)   & - & -  \\  
\hline
\end{tabular}
\end{center}
\label{tab:rawResults}
 \end{table}

\begin{table}[tb]
  \caption{
Summary of the transverse 
decay constant ($f_V^T(\mu)$) of the vector meson.
The scale is $\mu$=2 GeV.
}
\centering
\begin{tabular}{|c||c|c|c|} \hline
Ensemble   &  $a f_V^{T} / Z_T$ & $f_V^{T}(2 GeV)$ MeV 
& $\frac{f_V^{T}(2 GeV)}{f_V} $  \\ \hline
$B_1$  & 0.108(8)  & 194(15)  & 0.83(4) \\  
$B_2$  & 0.109(3)  & 195(5)   & 0.74(2) \\ 
$B_3$  & 0.111(2)  & 198(6)   & 0.75(1) \\ 
$B_4$  & 0.113(3)  & 203(6)   & 0.75(1) \\ 
$B_5$  & 0.128(5)  & 218(8)   & 0.78()  \\
$B_6$  & 0.109(5)  & 196(9)  & 0.73(2)   \\  
$C_1$  & 0.089(7)  & 214(18) & 0.77(5)   \\ 
$C_2$  & 0.081(4)  & 193(9)  & 0.71(2)   \\ 
$C_3$  & 0.089(3)  & 214(8)  & 0.79(2)   \\ 
$C_4$  & 0.090(3)  & 215(7)  & 0.76(1)   \\ 
\hline
\end{tabular}
\label{tab:rawResultstrans}
\end{table}

\begin{figure}
\begin{center}
\includegraphics[scale=0.5,angle=270,clip]{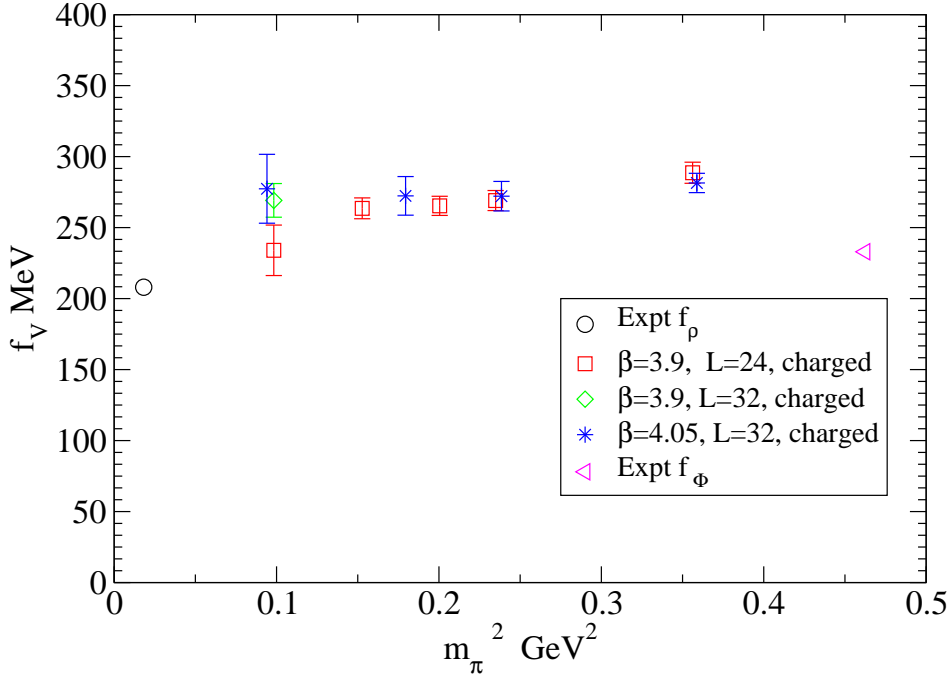}
\end{center}
\caption {
The leptonic decay constant of the vector meson 
(as defined in equation~\ref{eq:rhodecayDEFN})
is plotted 
as a function of the square of the pion mass.
The experimental points for the $\rho$ 
and $\phi$ are also included.
}
\label{fig:rhodecay}
\end{figure}

 The results for the leptonic decay constant are reported in
table~\ref{tab:rawResults} and the results for the transverse $\rho$
decay constant are in table~\ref{tab:rawResultstrans}. 
The decay constants from the neutral and charged vector mesons
agree within the errors. We now only consider the decay constants
of charged vector mesons.
In
figure~\ref{fig:rhodecay} we plot the decay constant of the vector meson
as a function of the square of the pion mass.
There is reasonable scaling between the decay
constants at $\beta = 3.9$ and $\beta = 4.05$.
The data with larger
masses also disagree with the value of the decay constant of the  $\phi$ meson. The $\phi$ has
a small decay width (4.26(4) MeV), so we might expect to be able get the
properties of this meson correctly.
 However, the $\phi$ is considered to be mostly 
$\overline{s} \gamma_\mu s$, so our neglect of the dynamics of the
strange quark could be important.

It has been found that chiral perturbation theory is required to
extrapolate the decay constants of the light pseudoscalar mesons to
their values at the physical quark
masses~\cite{Leutwyler:2008ma,Necco:2007pr}.  As discussed in
section~\ref{se:Sresults} the application of effective Lagrangian
techniques to study the $\rho$ meson is problematic because of the large
mass of the $\rho$ meson relative to the chiral
scale~\cite{Bruns:2004tj}. There are expressions for quark mass
dependence of the vector meson decay constants 
in~\cite{Bijnens:1998di}. The corrections due to loops start at $m_q \log
m_q$ and $m_q^{3/2}$. Given the size of the statistical errors on the
decay constants we didn't try to include any chiral corrections in the
chiral extrapolations.
A simple fit, linear in the square of the pion mass, 
of the $\beta=3.9$ data gives $f_{\rho}^{phys}$ = 239(18) MeV
and $f_{\phi}^{phys}$ = 308(29) MeV.

At the moment there are no results for the mass dependence of the
transverse leptonic decay  constants from effective field theory,
however the  formalism for tensor currents has  started to be
developed~\cite{Mateu:2007ha,Cata:2007ns}. It will be interesting to see
the predictions for the  mass dependence of the ratio of the transverse
to leptonic decay constant from effective field theory, because this
will  test whether a chiral extrapolation of the ratio  of the leptonic
to transverse decay constant results in a cancellation of systematic
errors as is hoped.

There has not been a definitive unquenched calculation of the  leptonic
decay constant of the $\rho$ meson, although there have been many
attempts. Lewis and Woloshyn came within  1\% of the experimental
result for the  $f_\rho$ in a quenched QCD calculation using the D234
improved action~\cite{Lewis:1996qv}. Lewis and Woloshyn summarise older
quenched calculations~\cite{Lewis:1996qv}. 
SESAM reported leptonic decay constants for vector mesons 
that agreed with experiment at the 20\% level from an unquenched
lattice QCD calculation with Wilson fermions~\cite{Eicker:1998sy}.
CP-PACS~\cite{AliKhan:2001tx}
from unquenched calculations with the tadpole improved clover action
found that they couldn't do a reliable continuum extrapolation of
$f_\rho$. CP-PACS~\cite{AliKhan:2001tx} found the non-perturbative and
perturbative renormalisation factors  to be very different. QCDSF
obtained  $f_\rho$ = 256(9) MeV from an unquenched calculation with
clover fermions~\cite{Gockeler:2005mh}. Hashimoto and
Izubuchi~\cite{Hashimoto:2008xg} obtained $f_{\rho}$ = 210(15)
MeV from a $n_f$ = 2 unquenched calculations that use domain wall
fermions. However this calculation also found that $r_0^{phys}$ =
0.549(9) fm from the mass of the $\rho$ meson, so we expect that this is
the reason for obtaining a number close to the physical point.

\begin{figure}
\begin{center}
\includegraphics[scale=0.5,angle=270,clip]{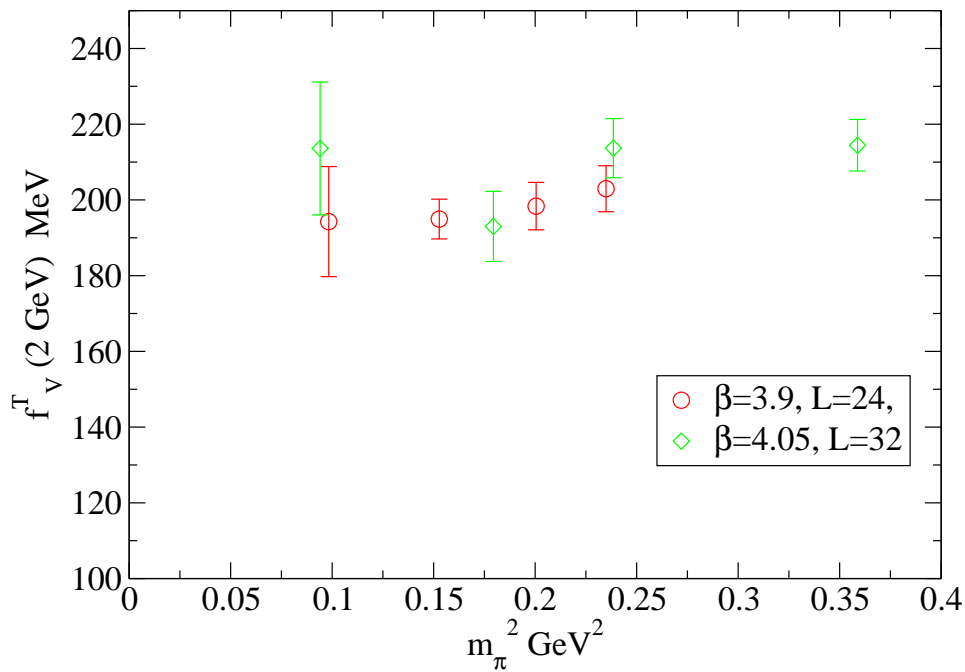}
\end{center}
\caption {
The transverse decay constant of the vector meson 
is plotted 
as a function of the square of the pion mass.
}
\label{fig:rhodecayTRANS}
 \end{figure}

\begin{figure}
\begin{center}
\includegraphics[scale=0.5,angle=270,clip]{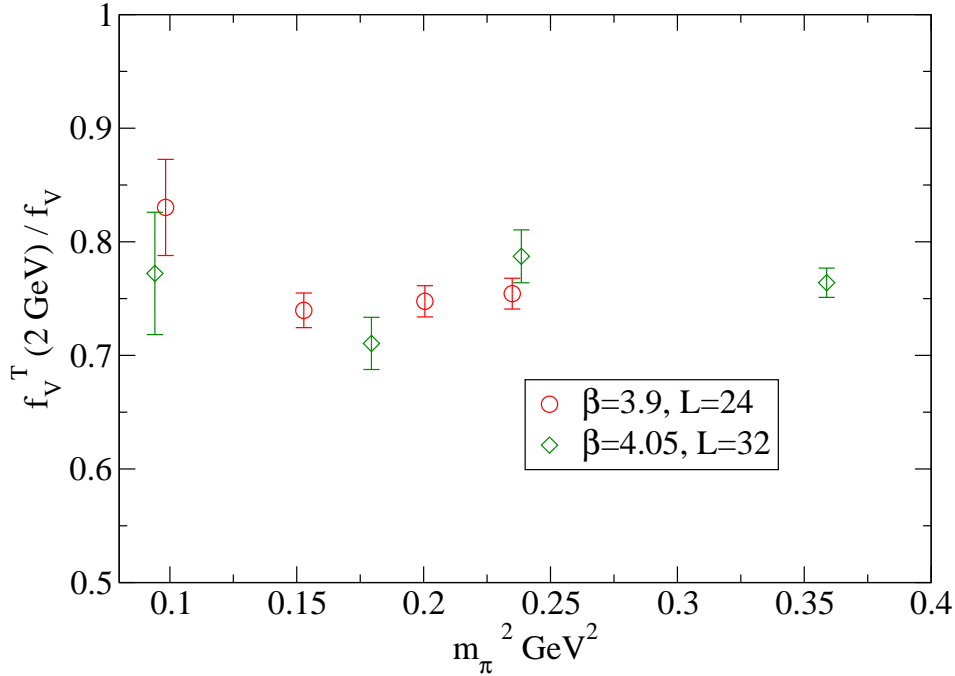}
\end{center}
\caption {
The ratio of transverse to leptonic decay
constant of the $\rho$ meson as a function of the square
of the pion mass. The transverse decay constant is at the 
scale of 2 GeV.
}
\label{fig:rhodecayRATIO}
\end{figure}

In figure~\ref{fig:rhodecayTRANS} we plot the transverse decay
constant of the vector meson as a function of the pion mass
squared in physical units. 
The ratio of transverse to leptonic decay constant is 
plotted in figure~\ref{fig:rhodecayRATIO}.

\begin{table}[tb]
  \caption{
Summary of results for transverse decay constants
of the $\rho$ and $\phi$ meson. 
We only include the result from the finest lattice of
Braun et al.~\cite{Braun:2003jg}.
}
\begin{center}
\begin{tabular}{|c|c|c|c|c|} \hline
Group & Method & $f_\rho^T$ (2 GeV)  &   $f_\phi^T$ (2 GeV) 
& $\frac{f_\rho^T}{f_\rho}$
\\ \hline
Ball et al.\cite{Ball:1998kk,Ball:1996tb,Bakulev:1999gf} & sum rule & 
155(10) & 208(15) & 0.74(3)  \\
Becirevic et al.~\cite{Becirevic:2004qv} & quenched lattice & 
150(5) & 177(2) & $0.72(2)^{+2}_{0}$ \\
Braun et al.~\cite{Braun:2003jg} & quenched lattice & 
154(5) & 182(2) & 0.74(1) \\
QCDSF \cite{Capitani:1999zd} & quenched lattice & 
149(9) & - & - \\
QCDSF \cite{Gockeler:2005mh} & unquenched lattice & 
168(3) & - & - \\
RBC-UKQCD~\cite{Allton:2008pn} & unquenched lattice  & 143(6)  & 175(2)
& 0.69(3)  \\
\hline
This work & unquenched lattice  & 184(15) & - & - \\
This work (ratio method) & unquenched lattice  &  159(8) &  -  & 0.76(4) \\
\hline
\end{tabular}
\end{center}
\label{tab:otherResults}
\end{table}

A collection of results for the transverse decay constants are presented
in table ~\ref{tab:otherResults}. We also present results from using the
ratio of tensor to vector correlators in the bootstrap analysis, that we
call the ``ratio method''. In~\cite{Dimopoulos:2008hb} the ETM
collaboration  presents results for
$\frac{f_{K^{\star}}^T}{f_{K^{\star}}}$ in a partially quenched analysis
on the same configurations. We see that our result for $f_\rho^T$ (2
GeV) is approximately 30 MeV higher than most previous  results. The
RBC-UKQCD collaboration also report a result for the transverse decay
constant of the $K^{\star}$ meson. Only QCDSF~\cite{Gockeler:2005mh} 
compute  $f_\rho^T$ on its own, all the others compute
$\frac{f_\rho^T}{f_\rho}$ and then multiply by experiment  value for
$f_\rho$. 

From what we call the unitary $\phi$ analysis we obtain $f_\phi^{T}$ =
170(14) MeV from the ratio method and  $f_\phi^{T}$ = 222(26) MeV from
the direct method, both at the scale of 2 GeV. 
These can be compared with the results in 
table~\ref{tab:otherResults}.

Cata and  Mateu~\cite{Cata:2008zc} (see also~\cite{Chizhov:2003qy})
have argued that in the  large $N_c$
limit that $\frac{f_\rho^T}{f_\rho}$  = $\frac{1}{\sqrt{2}}$. Their
result is consistent with  the lattice results in
table~\ref{tab:otherResults} for both the quenched and unquenched
results. 
There is some ambiguity in the large $N_c$ result, because it
doesn't depend on the renormalisation scale as it should do. There are
also predictions for the tensor decay constants of the excited vector
mesons from large $N_c$~\cite{Cata:2008zc}, that in principle could be
measured in future lattice QCD calculations that use modern variational
techniques~\cite{McNeile:2006qy}.

\section{Testing the mass dependence of the KRSF relations}

 \begin{figure}
\begin{center}
\includegraphics[scale=0.5,angle=270,clip]{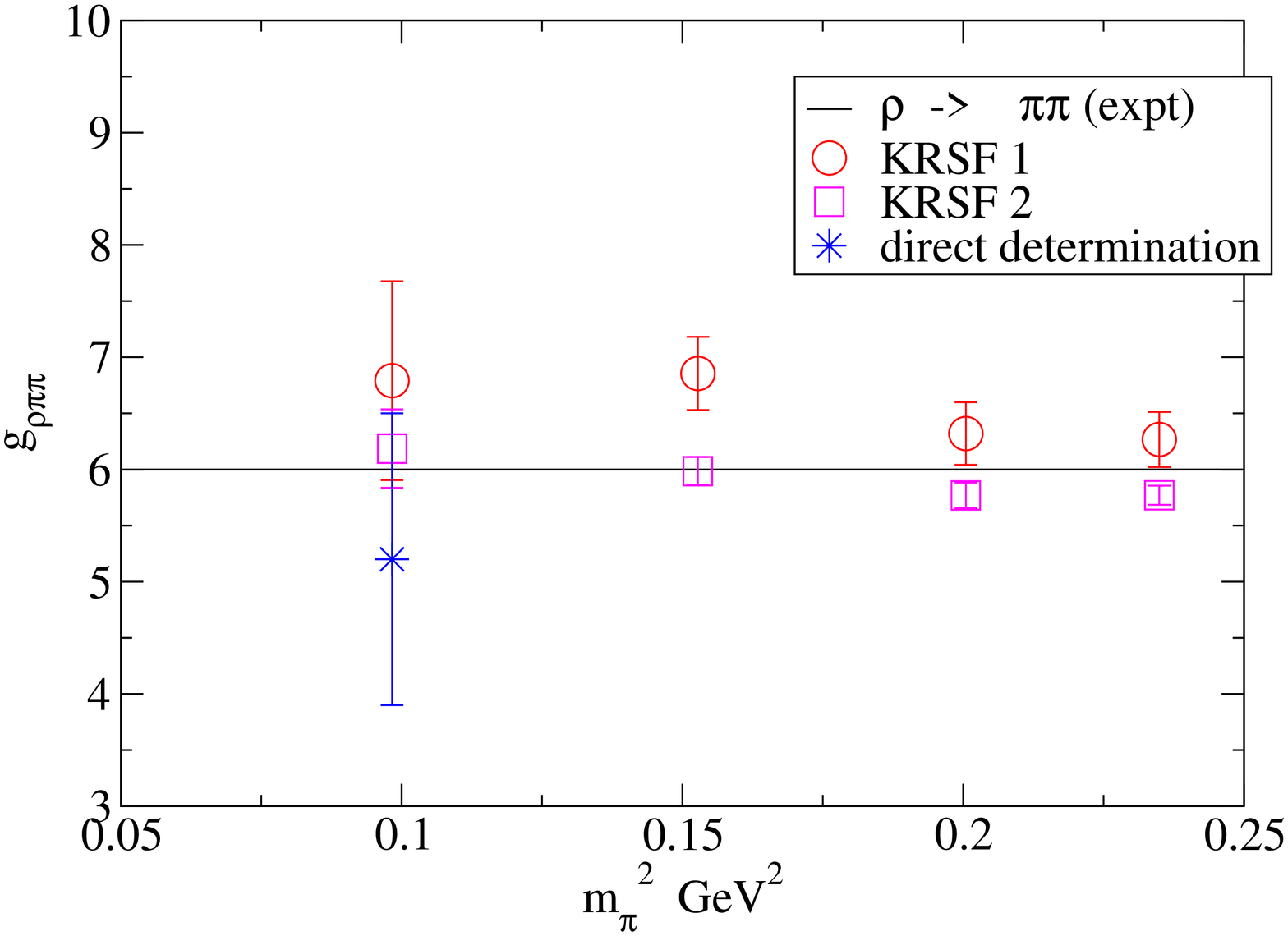}
\end{center}
\caption {
Comparing the $g_{\rho\pi\pi}$ coupling from the two KRSF relations
with experiment and the direct determination described in 
section~\ref{se:Sresults}.
}
\label{fig:krsfplot}
\end{figure}

In this paper we have discussed the mass of the 
$\rho$ meson, the lepton decay constant $f_\rho$,
and the coupling $g_{\rho\pi\pi}$ for $\rho$ decay to 
$\pi\pi$. Perhaps surprisingly there are postulated connections between the 
three constants, that are called the KRSF 
relationships~\cite{Kawarabayashi:1966kd,Riazuddin:1966sw}.
The original derivation of the KRSF relations used 
the application of the PCAC relation to $\rho$ 
decay~\cite{Kawarabayashi:1966kd,Riazuddin:1966sw}. However, the KRSF
relations are also predictions of some effective field theories
of mesons (see Birse for a review~\cite{Birse:1996hd}), 
such as those with ``hidden symmetry''~\cite{Bando:1987br}
and the vector realisation of chiral symmetry~\cite{Georgi:1989xy}.

Equation~\ref{eq:krsf1} and equation~\ref{eq:krsf2} 
are known as the KRSF1 and KRSF2 relationships~\cite{Bando:1987br}
respectively.
\begin{equation}
f_\rho \frac{m_\rho}{\sqrt{2}} = f_\pi^2 g_{\rho \pi\pi}  
\label{eq:krsf1}
\end{equation}
\begin{equation}
m_\rho^2 = f_\pi^2 g_{\rho \pi\pi}^2  
\label{eq:krsf2}
\end{equation}
We are using the convention where the physical 
pion decay constant is $f_\pi$  = 130.7 MeV.
In the effective field theory written down 
by Georgi~\cite{Georgi:1989xy}, there is an additional
2 on the right hand size of equation~\ref{eq:krsf2} that
makes his model not agree with experiment very well.
The KSRF relations can also be analyzed using
AdS/CFT~\cite{Son:2003et,Hong:2004sa}.

In figure~\ref{fig:krsfplot} we plot $g_{\rho \pi\pi}$
from equation~\ref{eq:krsf1} and equation~\ref{eq:krsf2} 
for the $\beta=3.9$ data, as a function of the 
square of pseudoscalar meson. With in the size
of the error bars, the value of  $g_{\rho \pi\pi}$
is relatively independent of the pseudoscalar mass.
This will be a useful test for hadronic effective field theories
that include the $\rho$ meson.

Some of the work on the KRSF relation in effective field theory is
used as a qualitative guide to building technicolor models of
electroweak symmetry breaking~\cite{Bando:1987br,Georgi:1989xy}.

\section{Conclusions} \label{SE:conclusions}

The first publication from the ETM collaboration showed 
impressive agreement between the predictions of chiral
perturbation theory and the lattice results~\cite{Boucaud:2007uk}.
In this paper we have found that getting agreement between the 
lattice results and the experimental data for the $\rho$, $b_1$, $a_0$
mesons is much harder. The statistical errors on the masses and couplings
are too large to look for subtle effects in the chiral extrapolation
models. 
More work, using the variational basis for the vector, the $b_1$ and
the $a_0$ states and more statistics will be needed to eventually test
the various chiral extrapolations.

We have started to explore using various tools, such as the
Adelaide~\cite{Leinweber:2001ac,Allton:2005fb,Armour:2005mk} method,
computation of decay widths, and looking for avoided level crossings,
to study resonance mesons on the lattice.
 Eventually, the issue of dealing with resonances in lattice QCD
will use L\"{u}scher's technique~\cite{Luscher:1991cf} and variants
of~\cite{Bernard:2008ax}. 
These methods
are computationally intensive, so more pragmatic approaches to studying strong
decays on the lattice are still important at this time.  L\"{u}scher's
technique for resonances was recently applied to the $\rho$ meson by
the CP-PACS collaboration~\cite{Aoki:2007rd}.

\section{Acknowledgments}

We thank Jozef Dudek for discussions about chiral extrapolations
and open decays. 
We thank all members of ETMC for a very fruitful collaboration and, in
particular, Giancarlo Rossi, P. Dimopoulos, 
Xu Feng, G. Herdoiza and S. Simula for useful comments on the paper.
We acknowledge the use of the NW grid for part of this analysis.



\end{document}